\documentstyle[12pt]{article}
\begin{document}
\thispagestyle{empty}
\begin{center}
\LARGE \tt \bf{On the stability of de Sitter inflationary background solution in scalar-tensor cosmology}
\end{center}
\vspace{2.5cm}
\begin{center} {\large By  L.C. Garcia de Andrade\footnote{Departamento de
F\'{\i}sica Te\'{o}rica - Instituto de F\'{\i}sica - UERJ
Rua S\~{a}o Fco. Xavier 524, Rio de Janeiro, RJ
Maracan\~{a}, CEP:20550-003 , Brasil.E-mail:garcia@dft.if.uerj.br.}}
\end{center}
\vspace{2.0cm}
\begin{abstract}
The stability of de Sitter metric in spacetimes where inflaton fields are non-minimally coupled to propagating torsion is investigated. It is shown that the de Sitter background metric is unstable against small perturbations. Axionic torsion is generated by inflaton fluctuations and the analysis of cosmological perturbations in Einstein-Cartan-Klein-Gordon (ECKG) equations is undertaken. Instabilities in the inflaton massless modes are found when the minimal coupling constant dominates. Inflaton fluctuations trigger axionic torsion  vector fluctuations. The ECKG equations reduce to Einstein-de Sitter equations in the case of a constant background inflaton field.
\end{abstract}
\newpage
\pagestyle{myheadings}
\markright{\underline{On the stability of de Sitter metric and torsion.}}
\section{Introduction}  

   The interest in cosmological density perturbations \cite{1} has grown recently with the new MAP and other recently experiments proposed to measured the CMBR. Examples of density and metric perturbations which are correlated problems, have been put forward recently specially by Fabris and his group \cite{2} which computed the density and metric  perturbations in string like fluids in scalar field gravity and Maroto and Shapiro \cite{3} which computed the stability of de Sitter metric in the case of dilatonic solutions in higher-order gravity and arrived at the conclusion that the metric is unstable. Fabris et al have also found gravitational instabilities in the longwavelength limit. In this letter we show that the stability of the metric in spacetimes with torsion with non-minimal coupled inflatons is by no means absolute but depends on the region of the spacetime and boundary conditions. A similar situation has also appeared before as was noted by Silk \cite{4} while investigating the stability of rotating universes with G\"{o}del metric showing that the metric was stable along one direction and unstable along the other. This situation also appeared in Fabris et al considered the cosmological perturbations of string-like fluids in hydrodynamical representations \cite {2} where density fluctuations of fluids with negative pressure exhibited decreasing time behaviour in the long wave limit but are strongly unstable in the small wave limit with the gravitational waves being well behaved. It is very important to call the attention to the fact that the need of non-minimal coupling between inflatons and Cartan torsion is fundamental since there is no minimal coupling between torsion and scalar fields as has been pointed out before by Hehl et al. \cite {5} and Stornaiolo in his PhD. thesis \cite{6}. This is due to the fact that torsion is a a half-integer field while scalars are spin-0 fields. This situation also happens between torsion and spin-1 fields like photons as in the case of the non-minimal coupling between massive photons and torsion \cite{7}. Non-minimal coupled behaviour can explain important issues of physics such as the value of primordial magnetic fields in the universe. Someone could ask why to bother use non-minimal coupled between torsion and inflatons to investigate inflationary cosmology with torsion if there are studies that show that it is where there are inflationary models with torsion but without inflatons as shown by D.Palle \cite{8}. The answer to this question is simple and considers the fact that most modern theories of inflation that has been considered so far \cite{9} have dealt with inflaton fields. The present paper is organized as follows. In section 2 we show that the Einstein-de Sitter equation is obtained as a constant inflaton field reduction from the ECKG field equations and show that the inflaton fluctuations may generate axionic torsion vector fields. In section 3 we show that the inflaton fluctuations also generate torsion fluctuations and metric fluctuations around the background de Sitter solution and show that the solution is unstable exactly in the string inspired model of Maroto and Shapiro.In this last section we also discuss the case of massless inflatons. 
\section{De Sitter ground state in ECKG cosmology}
 Let us now consider the non-minimal coupled Lagrangean density given by
\begin{equation}
L=\frac{1}{2}\sqrt{-g}[\frac{1}{8{\pi}G}R+g^{ab}{\partial}_{a}{\phi}{\partial}_{b}{\phi}-(m^{2}+{\epsilon}R){\phi}^{2}]
\label{1}
\end{equation}
where ${a,b=0,1,2,3}$ and $R({\Gamma})$ is the Riemann-Cartan spacetime Ricci scalar curvature which interacts with the inflaton field ${\phi}$ through the last term.Variation of the action $S=\int{d^{4}x L}$ w.r.t. to the scalar field ,contortion tensor and metric yield the following field equations 
\begin{equation}
g^{ab}{\partial}_{a}{\partial}_{b}{\phi}-g^{ab}{{\Gamma}^{*}}^{c}_{ab}{\partial}_{c}{\phi}+(m^{2}+{\epsilon}R){\phi}=0
\label{2}
\end{equation}
and
\begin{equation}
T_{a}={T_{ab}}^{b}=24{\pi}G{\epsilon}(1-8{\pi}G{\epsilon}{\phi}^{2})^{-1}{\partial}_{a}{\phi}^{2}
\label{3}
\end{equation}
which as shown by Stornaiolo the only surving part of the torsion tensor is the torsion vector. The final equation is the Einstein-Cartan (EC) type equation which however now possess a propagating torsion structure and torsion is dynamical contrary to the more usual EC gravity.Here ${\Gamma}^{*}$ is the Riemannian connection. All quantities with an upper asterix will denote a Riemannian quantity. The ECKG type field equation reads

\begin{eqnarray}
{G_{ab}}^{*}=-6(16{\pi}G{\epsilon})^{2}(1-8{\pi}G{\epsilon}{\phi})^{-2}{\phi}^{2}[2{\partial}_{a}{\phi}{\partial}_{b}{\phi}-g_{ab}g^{cd}{\partial}_{c}{\phi}{\partial}_{d}{\phi}]\nonumber \\
-(1-8{\pi}G{\epsilon}{\phi})^{-1}[(1-2{\epsilon}){\partial}_{a}{\phi}{\partial}_{b}{\phi}+(2{\epsilon}-\frac{1}{2})g_{ab}g^{cd}{\partial}_{a}{\phi}{\partial}_{b}{\phi}\nonumber \\
+\frac{1}{2}m^{2}{\phi}^{2}g_{ab}+2{\epsilon}g_{ab}{\phi}{\ddot{\phi}}-2{\epsilon}g_{ab}{\phi}g^{cd}{{\Gamma}^{*}}^{0}_{cd}\dot{\phi}\nonumber \\
-2{\epsilon}{\phi}{\partial}_{a}{\phi}{\partial}_{b}{\phi}+2{\epsilon}{\phi}{{\Gamma}^{*}}^{0}_{ab}\dot{\phi}]
\label{4}
\end{eqnarray}

It is easy to note that by considering the inflaton fluctuation around a constant background field $<{\phi}>={\phi}_{0}= constant$ such as 
\begin{equation}
{\phi}={\phi}_{0}+ {\delta}{\phi}
\label{5}
\end{equation}

at ${\phi}_{0}$ the ECKG field equations reduce to
\begin{equation}
{G_{ab}}^{*}={\Lambda} g_{ab}
\label{6}
\end{equation}
where ${\Lambda}$ is the Cosmological constant given here by 
\begin{equation}
{\Lambda}= \frac{1}{2}m^{2}{{\phi}_{0}}^{2}
\label{7}
\end{equation}
Therefore we note that in our model the cosmological constant vanishes in the case of background massless inflaton fields. This is not a necessary condition for inflation but it can be modified by a change of scale of the form ${\phi}_{0}= {\phi}'-{\phi}$. The proof here is important for the remaining sections since to perform fluctuations on de Sitter background metric we needed to prove that this metric was indeed a background solution of ECKG equations. Let us now consider the same inflaton fluctuation above and show that in first approximation (this is valid since we are considering small perturbations) they can generate axionic torsion fields. This can be easily done by expanding the expression (\ref{3}) as follows
\begin{equation}
T_{a}={T_{ab}}^{b}=24{\pi}G{\epsilon}(1-8{\pi}G{\epsilon}{({\phi}_{0}+{\delta}{\phi})}^{2})^{-1}{\partial}_{a}({{\phi}_{0}}^{2}+2{\phi}_{0}{\delta}{\phi}+{{\delta}{\phi}}^{2})
\label{8}
\end{equation}
By keeping terms till first order in ${\delta}{\phi}$ and ${\epsilon}$ one obtains the following expression
\begin{equation}
T_{a}=48{\pi}G{\epsilon}{\phi}_{0}{\partial}_{a}{\delta}{\phi}
\label{9}
\end{equation}
which shows that in the case of a constant background state ${\phi}_{0}$ Cartan torsion vanishes and we are left with a torsion-free Einstein-de Sitter gravity. The idea of considering Einstein-Cartan gravity as a fluctuation around general relativity can be better understood if one considers that the Cartan cortortion can be expressed as a fluctuation of connections in the expression
\begin{equation}
{K^{a}}_{bc}={\delta}{\Gamma}^{a}_{bc}={\Gamma}^{a}_{bc} - {{\Gamma}^{*}}^{a}_{bc}
\label{10}
\end{equation}
where ${\Gamma}^{a}_{bc}$ is the Riemann-Cartan connection.
\section{Inflaton and metric perturbations}
Note that the equation for the inflaton potential equation can be obtained from the above equation (\ref{2}) when one considers that the inflaton field only depends on time
\begin{equation}
\ddot{\phi}-2H_{0}\dot{\phi}+(2m^{2}+{\epsilon}R^{*}){\phi}=0
\label{11}
\end{equation}
where we are considering the Riemannian Ricci scalar since we are assuming that the background geometry is Riemannian before the fluctuation.We also assume that the background geometry obeys the de Sitter metric
\begin{equation}
ds^{2}=dt^{2}-e^{2H_{0}t}(dx^{2}+dy^{2}+dz^{2})
\label{12}
\end{equation}
where $H_{0}$ is the Hubble constant. Perturbation of the equation (\ref{11}) yields
\begin{equation}
{\delta}{\ddot{\phi}}-2H_{0}{\delta}{\dot{\phi}}+(2m^{2}+R^{*}){\delta}{\phi}+{{\delta}R^{*}}{\phi}=0
\label{13}
\end{equation}
since for the de Sitter metric the Ricci scalar is $R^{*}=-6H_{0}^{2}$ equation (\ref{13}) reduces to
\begin{equation}
{\delta}{\ddot{\phi}}-2H_{0}{\delta}{\dot{\phi}}+2(m^{2}-6{\epsilon}H_{0}^{2}){\delta}{\phi}=0
\label{14}
\end{equation}
the solution of this equation is rather simple if take the slow roll approximation of inflationary cosmology and drop the second time derivatives on ${\phi}$ which yields
\begin{equation}
{\delta}{\dot{\phi}}+(6{\epsilon}H_{0}^{2}-m^{2}){\delta}{\phi}=0
\label{15}
\end{equation}
by considering the constraint ${6{\epsilon}H_{0}^{2}-m^{2}}>0$
one obtains the following solution for the inflaton fluctuation
\begin{equation}
{\delta}{\phi}=Ae^{-\frac{6{\epsilon}H_{0}^{2}-m^{2}}{H_{0}}t}
\label{16}
\end{equation}
which shows that this perturbation is damped with time.Here A is an integration constant that should be constraint to be very small since we are dealing with small cosmological perturbations.From the expression of torsion above we can compute the fluctuation on the torsion zero component which is the only surviving component of the torsion vector in terms of inflaton fluctuation given in (\ref{10}) as
\begin{equation}
{\delta}T_{0}=-\frac{6\dot{\phi}{\delta}{\dot{\phi}}}{{\phi}^{2}}
\label{17}
\end{equation}
where to simplify matters we consider the approximation of high values inflaton fields.The inflaton field and its derivative can be obtained by solving the equation (\ref{12}) which in the slow roll approximation yields 
\begin{equation}
{\phi}=Be^{-\frac{6{\epsilon}H_{0}^{2}-m^{2}}{H_{0}}t}
\label{18}
\end{equation}
where the integration constant B now need not be necessarily small.Time derivative of this expression reads
\begin{equation}
{\dot{\phi}}=\frac{6{\epsilon}H_{0}^{2}-m^{2}}{H_{0}}{\phi}
\label{19}
\end{equation}
Substitution of (\ref{12}) and (\ref{13}) into the expression (\ref{11}) yields
\begin{equation}
{\delta}T_{0}=-6A^{2}\frac{6{\epsilon}H_{0}^{2}-m^{2}}{H_{0}}
\label{20}
\end{equation}
which shows that in this approximation the torsion fluctuation is constant.Remains now to investigate the fluctuation of the last EC type equation to study the stability of de Sitter metric in the case of non-minimal coupling between torsion and inflaton fields.Let us now consider the fluctuation on the $(00)$ component of the Einstein tensor as
\begin{equation}
{\delta{G^{*}_{00}}}=-\frac{6{\epsilon}H_{0}^{2}-m^{2}{\phi}}{{\epsilon}{\phi}^{2}}{\delta{\phi}}
\label{21}
\end{equation}
Knowing that 
\begin{equation}
{\delta{G^{*}_{00}}}=\frac{1}{4}(\ddot{h}+2H_{0}\dot{h})
\label{22}
\end{equation}
where $h_{ab}={\delta}g_{ab}$ is the metric perturbation. Here 
\begin{equation}
{\delta{G^{*}_{00}}}= 4\dot{\phi}\dot{\delta{\phi}}
\label{23}
\end{equation}
Thus from these last equations one obtains
\begin{equation}
(\ddot{h}+2H_{0}\dot{h})= 4\dot{\phi}\dot{\delta{\phi}}
\label{24}
\end{equation}
One obtains an equation for the de Sitter metric fluctuation in terms of the inflaton field fluctuation.This equation can be expressed also in terms of torsion vector fluctuations as
\begin{equation}
\ddot{h}+2H_{0}\dot{h} = \frac{2}{3}{\phi}^{2}{\delta}T_{0}
\label{25}
\end{equation}
Here $h_{kk}=ha^{2}$ where ${k=1,2,3}$ and where we have used the gauge $h_{a0}=0$. Before we solve these equation we need to get the inflaton solution since we have a coupled inflaton-torsion-metric perturbation system. Thus by considering the case of masless inflatons in the above complete inflaton fluctuation equation we obtain
\begin{equation}
{\phi}= e^{{\alpha}t}
\label{26}
\end{equation}
where ${\alpha}_{\pm}= H_{0}(1 {\pm}\sqrt{1+12{\epsilon}})$. This reduces equation (\ref{25}) to 
\begin{equation}
\ddot{h}+2H_{0}\dot{h} = 4H_{0}(1 {\pm}\sqrt{1+12{\epsilon}}) 
\label{27}
\end{equation}
Now we have two cases. The first is where the non-minimal coupling ${\epsilon}$ vanishes. In this case we have the following perturbations modes for de Sitter metric
\begin{equation}
h= e^{-2H_{0}t}+2t
\label{28}
\end{equation}
In the case ${\epsilon}$ goes to infinity or the non-minimal coupling dominates is given by
\begin{equation}
h= e^{-2H_{0}t}+ 4\sqrt{3{\epsilon}}t
\label{29}
\end{equation}
which is unstable as well.
The inflaton is possess two modes one stable and one unstable which are given by

\begin{equation}
{\delta}{\phi}_{\pm}= e^{\pm{H_{0}\sqrt{12{\epsilon}}}t}
\label{30}
\end{equation}
which is also unstable. Therefore one may say that in both cases de Sitter background solution is unstable as in Maroto-Shapiro case where the instability of de Sitter solution depends on parametrization but once parametrization is choosed one may always show that the de Sitter metric is unstable. A more detailed account of the problems discussed here may appear elsewhere include the computation of density perturbations for this non-minimally coupled theory of gravity with propagating torsion. 
\section*{Acknowledgement}
I am very much indebt to Prof. P.S.Letelier and Prof. I.L.Shapiro for helpful discussions on the subject of this  paper. Financial support from CNPq. is gratefully ackowledged.

\end{document}